\documentclass[epj]{svjour}
\usepackage{graphics}
\usepackage{amssymb}
\begin{document}
\title{ DELTA EXCITATION IN DEUTERON-PROTON ELASTIC
SCATTERING}
\author{N.B.Ladygina\inst{1}% etc
\thanks{\emph{Present address:} nladygina@jinr.ru}%
}                     % Do not remove
\offprints{}          % Insert a name or remove this line
\institute{Joint Institute for Nuclear Research, LHEP, Dubna, Russia}
\date{Received: date / Revised version: date}
\abstract{
 Deuteron-proton elastic scattering is studied in the multiple
 scattering expansion formalism. 
The four contributions are taken into account: one-nucleon-exchange,
single- and double scattering, and $\Delta$-isobar excitation.
The presented approach was applied to describe the 
differential cross sections at deuteron energies between 
500 and 1300 MeV in a whole angular range. 
The obtained results are compared with the experimental data.
 % end of PACS codes
} %end of abstract
\maketitle
\section{Introduction}
In recent years heavy ion physics at high energies  has been very popular.
There are many efforts to study some systems containing a lot of nucleons.
However, the energy in 
such reactions is quite  large to appear the multiple nucleon scattering.
Therefore, the problem of the nucleon-nucleon interaction is topical up to now.
Even at low energies all the theoretical calculations are based on the 
phenomenological nucleon-nucleon potential. The problem becomes more
difficult when the energy is enough for a manifestation of inelastic channels. 

Elastic deuteron-proton scattering  is the simplest 
example of the hadron nucleus collision, because a deuteron is the simplest nucleus
containing only one proton and one neutron.
The study of the deuteron-proton elastic scattering has a longtime
story. The first nucleon-deuteron experiments were performed already in fifties
of the previous century \cite{sham}-\cite{postma}. Differential cross sections \cite{sham}-\cite{crewe}
and polarization  \cite{marshal}-\cite{postma}
were measured at a few hundred MeV. This topic has been intensively investigated in 
the 1970's and 80's \cite{igo1}-\cite{ghaz}.
Nowadays this reaction is still the subject of the investigations
\cite{cr500}-\cite{pkur}.

In the previous papers \cite{japh},\cite{epja} we considered this reaction in the multiple
 scattering expansion formalism. Three contributions, one-nucleon-exchange (ONE), single- scattering, and
double scattering, were taken into account.
 We got a reasonable agreement for the differential cross sections
 between the theoretical predictions
and the experimental data almost in a whole angular range. However, the rise of the differential cross sections
at the scattering angle larger than $140^\circ$ was not described in the approach.

In 1969 A.Kerman and L.Kisslinger  supposed that resonances can play an important role in the deuteron-proton
backward elastic scattering \cite{ker}. They suggested that a deuteron contains the $N^*$ resonance with
$0.5 - 1 \%$ probability. However, an analysis of the experimental data using the Kerman- Kisslinger model \cite{igo2}
has demonstrated strong dependence of the obtained results on a deuteron wave function.

 Later the double-scattering mechanism  with $\Delta$-isobar in the intermediate 
state was taken into account in the dp backward scattering. The significant
contribution of this diagram to the reaction was shown  in refs. \cite{kon26}-\cite{kon33}.
 However,  the double scattering with nucleon in an
intermediate state was not considered in these papers .
Perhaps, it was the reason why the description of  the differential cross sections energy dependence was not good enough.
  
The effort to take the $\Delta$-isobar into account in order to describe dp-elastic scattering was also done in  \cite{chiev}, \cite{deltuva}.
\cite{maeda}, where the $\Delta$-isobar term was included into the CD Bonn potential.
In these papers deuteron-proton scattering was considered in a whole angular range, not only at $\theta^*=180^\circ$.
Unfortunately, the process was
studied at low energies, $T_p\le 250$ MeV, where the $\Delta$-isobar  effects are negligible. 

In this paper we keep on considering  deuteron-proton elastic scattering in the framework
offered in  papers \cite{japh}, \cite{epja}. Here we add  the $\Delta$-isobar contribution in our approach. 
The next part is devoted to the description of the general theoretical model. We briefly
give  the basic  points of the multiple scattering approach.   In section 3 the $\Delta$-excitation term is considered in a detail.
Also the kinematics of the double scattering is given, both for the delta and nucleon in the intermediate state.
 In the last section
we discuss the obtained results and compare them with the experimental data.

\section{General formalism}
 According to the three-body collision theory,
the amplitude of  deuteron-proton elastic scattering $\cal J$
is defined by the matrix element of the transition operator $U_{11}$:
\begin{eqnarray}
U_{dp\to dp}&=&\delta(E_d+E_p-E^\prime_d-E^\prime _p) {\cal J}=
\\
&&<1(23)|[1-P_{12}-P_{13}]U_{11}|1(23)>~~.
\nonumber
\end{eqnarray}
Here the state $|1(23)>$ corresponds to the configuration, 
when  nucleons 2 and 3
form the deuteron state and nucleon 1 is free. 
Appearance of the permutation operators for two nucleons $P_{ij}$
reflects the fact that the initial and final states are antisymmetric
due to an exchange of the two particles.

Following the Alt-Grassberger-Sandhas formalism \cite{ags}, \cite{zig} we write the transition operator
$U_{\beta\alpha}$ as:
\begin{eqnarray}
U_{\beta\alpha}=-(1-\delta_{\beta\alpha})(H_0-z)-\sum_{\delta\ne\alpha}U_{\beta\delta}g_0t_\delta.
\end{eqnarray}
This transition operator depends on the  potential through the channel transition-operator $t_\alpha$
 of the two-particle subsystems: 
\begin{eqnarray}
t_\alpha=V_\alpha-V_\alpha g_\alpha V_\alpha
\\
t_\alpha=V_\alpha-V_\alpha g_0 t_\alpha,
\nonumber
\end{eqnarray}
where $g_0$ is a free three-particle propagator and $g_\alpha$ is a two-particle resolvent in the three-particle space.
In such a way we get the following equations for the transition operators with the rearrangement:  
\begin{eqnarray}
U_{11}&=&~~~~~~~~t_2g_0U_{21}+t_3g_0U_{31}
\nonumber\\
U_{21}&=&g_0^{-1}+t_1g_0U_{11}+t_3g_0U_{31}
\\
U_{31}&=&g_0^{-1}+t_1g_0U_{11}+t_2g_0U_{21}~,
\nonumber
\end{eqnarray}
where $t_1=t(2,3)$,etc., is a t-matrix of the two-particle interaction. The indices $ij$ for the
transition operators $U_{ij}$ denote free particles $i$ and $j$ in the final
 and initial states, respectively.

Iterating these equations up to the second order terms of  $t_i$ we can present
the reaction amplitude as a sum of the following four contributions:
 one nucleon exchange,
single scattering and double scattering with the nucleon and delta in the intermediate state, --
\begin{eqnarray}
\label{contrib}
{\cal J}_{dp\to dp}&=&{\cal J}_{ONE}+{\cal J}_{SS}+{\cal J}_{DS}+{\cal J}_{\Delta}
\end{eqnarray}

The first contribution into the $dp$-elastic scattering amplitude ${\cal J}$
in Eq.(\ref {contrib}) is the  one nucleon exchange (ONE) term.
\begin{eqnarray}
{\cal J}_{ONE}=-2<1(23)|P_{12}g_0^{-1}|1(23)>
\end{eqnarray}
The corresponding diagram is presented in Fig.1a. Applying the
definitions of the wave function of a moving deuteron and three-
nucleon free propagator, we can write ONE amplitude in the following 
form:
\begin{eqnarray}
\label{one}
{\cal J}_{ONE}&=&-2_{1(23)}<\vec p^\prime m^\prime \tau^\prime; 
-\vec P_d {\cal M}_d^\prime 0|\Omega_d^\dagger (23) P_{12}
%\nonumber
\nonumber\\
&&(E_d+E_p-\hat K_1-\hat K_2-\hat K_3 +i\varepsilon )
\\
&&\Omega_d(23)|
\vec P_d {\cal M}_d 0;\vec p m \tau>_{1(23)}~~,
\nonumber
\end{eqnarray}
where $m$, $m^\prime$ are spin projections of the initial and final protons,
$\tau$, $\tau^\prime$ are their isospin projections, respectively.  
The kinetic-energy operator has a standard definition,
 $\hat K_i|\vec p_i>=\sqrt{m_N^2+\vec p_i\-^2}|\vec p_i>$ .

After a straightforward calculation we have the expression
for the ONE amplitude
\begin{eqnarray}
\label{ONE}
&&{\cal J}_{ONE}=-\frac{1}{2}(E_d-E_p-\sqrt{m_N^2+\vec p~^2-\vec P_d\-^2})
\cdot
\nonumber\\
&&<\vec p^\prime m^\prime;-\vec P_d {\cal M}_d^\prime|\Omega^\dagger_d(23)
\\
&&[1+(\mbox{\boldmath$\sigma_1\sigma_2$})]
\Omega_d(23)|
\vec P_d {\cal M}_d;\vec p m >~~,
\nonumber
\end{eqnarray}
where the definition of  the permutation
operator in spin space 
 $P_{12}(\sigma )=\frac{1}{2}[1+(\mbox{\boldmath$\sigma_1\sigma_2$})]$ has been applied.

All the calculations are performed in the deuteron Breit frame, where
the deuterons move  in opposite directions with equal momenta (Fig.1).
%\begin{eqnarray}
%E_d&=&E^\prime_d=\sqrt{M_d^2+\vec Q^2}, ~~~~~~~~~
%E_p=E^\prime_p=\sqrt{m^2+\vec p^2},
%\nonumber\\
%\nonumber\\
%&&
%\hspace{2cm}
%(\vec p\vec Q)=-\vec Q^2.
%\end{eqnarray}
It allows us to minimize  the relative momenta of the nucleons in both
deuterons. As a consequence, the non-relativistic deuteron wave function
can be applied in the energy range under consideration.

In the rest frame the non-relativistic wave function  of the deuteron 
depends only on one variable $\vec p_0$, which is the
 relative momentum of the  outgoing proton and neutron:
\begin{eqnarray}
\label{dwf0}
&&<\mu_p \mu_n|\Omega_d|{\cal M}_d>=
\frac{1}{\sqrt{4\pi}}<\mu_p \mu_n|\{ u(p_0)+
\\
&&\hspace*{2cm}
\frac{w(p_0)}{\sqrt 8}
[3(\mbox{\boldmath$\sigma_1$} \hat p_0)(\mbox{\boldmath$\sigma_2$}\hat p_0)-(
\mbox{\boldmath $\sigma_1$ $\sigma_2$}
)]
\}|{\cal M}_d>,
\nonumber
\end{eqnarray}
where $u(p_0)$ and $w(p_0)$ describe  $S$ and $D$ components of 
the deuteron wave function   \cite{B}, \cite{cd}, \cite{par}, $\hat p_0$ is 
a unit vector in $\vec p_0$ direction.

In order to get the wave function of the moving deuteron, it is necessary to apply
the Lorenz transformations for the kinematical variables and Wigner rotations for 
the spin states. This procedure has been expounded in a detail in ref.\cite{japh}.
The proton-neutron relative momenta    for the initial
 $\vec p_0$ and final $\vec p_0^\prime $ deuterons are expressed in relativistic kinematics as: 
\begin{eqnarray}
\vec p_0&=&\vec p +\vec P_d\left[ 1+\frac{E_n+E^*}{E_p+E_n+E^*}\right]
\\
\vec p_0^\prime &=&\vec p +\vec P_d\left[ 1-\frac{E_n+E^*}{E_p+E_n+E^*}\right]~~.
\nonumber
\end{eqnarray}
Here $E_n=\sqrt{m_N^2+\vec p~^2-\vec P_d\-^2}$ 
 and
\\
 $E^*=\sqrt{(E_p+E_n)^2-\vec P_d\-^2}/2$ 
are the struck neutron energy in the moving deuteron frame and rest deuteron
frame, respectively. Note, that $|\vec p_0|=|\vec p_0^\prime |$.

The other term in the $dp$-elastic scattering amplitude 
Eq.(\ref {contrib}) is the single scattering (SS) one. 
\begin{eqnarray}
{\cal J}_{SS}&=&2<1(23)|[1-P_{12}]t_3|1(23)>
%\\
%{\cal J}_{DS}&=&2<1(23)[1-P_{12}]|t_3 g_0 t_2[1-P_{13}]|1(23)>~~,
\end{eqnarray}

The corresponding diagram is presented in Fig.(1b).
Following a  standard procedure we get the expression 
for the single scattering  amplitude:
\begin{eqnarray}
\label{ss}
{\cal J}_{SS}&=&\int d\vec p_2 <-\vec P_d {\cal M}_d^\prime |
\Omega_d^\dagger|
\vec p_2 m^{\prime\prime}, -\vec P_d-\vec p_2 m_3^\prime>
\nonumber\\
&&\hspace{-1cm}
<\vec p~^\prime m^\prime, -\vec P_d-\vec p_2 |
\frac{3}{2}t^1_{12}+\frac{1}{2}t^0_{12}|
\vec p m, \vec P_d -\vec p_2 m_2^\prime >
\nonumber\\
&&<\vec p_2 m^{\prime\prime},\vec P_d-\vec p_2 m_2^\prime|
\Omega_d|\vec P_d {\cal M}_d>~~.
\end{eqnarray}
The relative momenta of the two nucleons for the initial and final 
deuterons are
\begin{eqnarray}
\vec p_0&=&\vec p_2 -\vec P_d\frac{E_2+E^*}
{E_2+E_3+2E^*}
\\
\vec p_0^\prime &=&\vec p_2 +\vec P_d
\frac{E_2+E^{\prime *}}
{E_2+E_3^\prime +2E^{\prime *}}~~,
\nonumber
\end{eqnarray}
where  nucleons energies $E_2$, $E_3$, $E_3^\prime $ 
in the reference frame are defined by a 
standard manner (Fig.1b)
\begin{figure*}[t]
\resizebox{1\textwidth}{!}{
\includegraphics{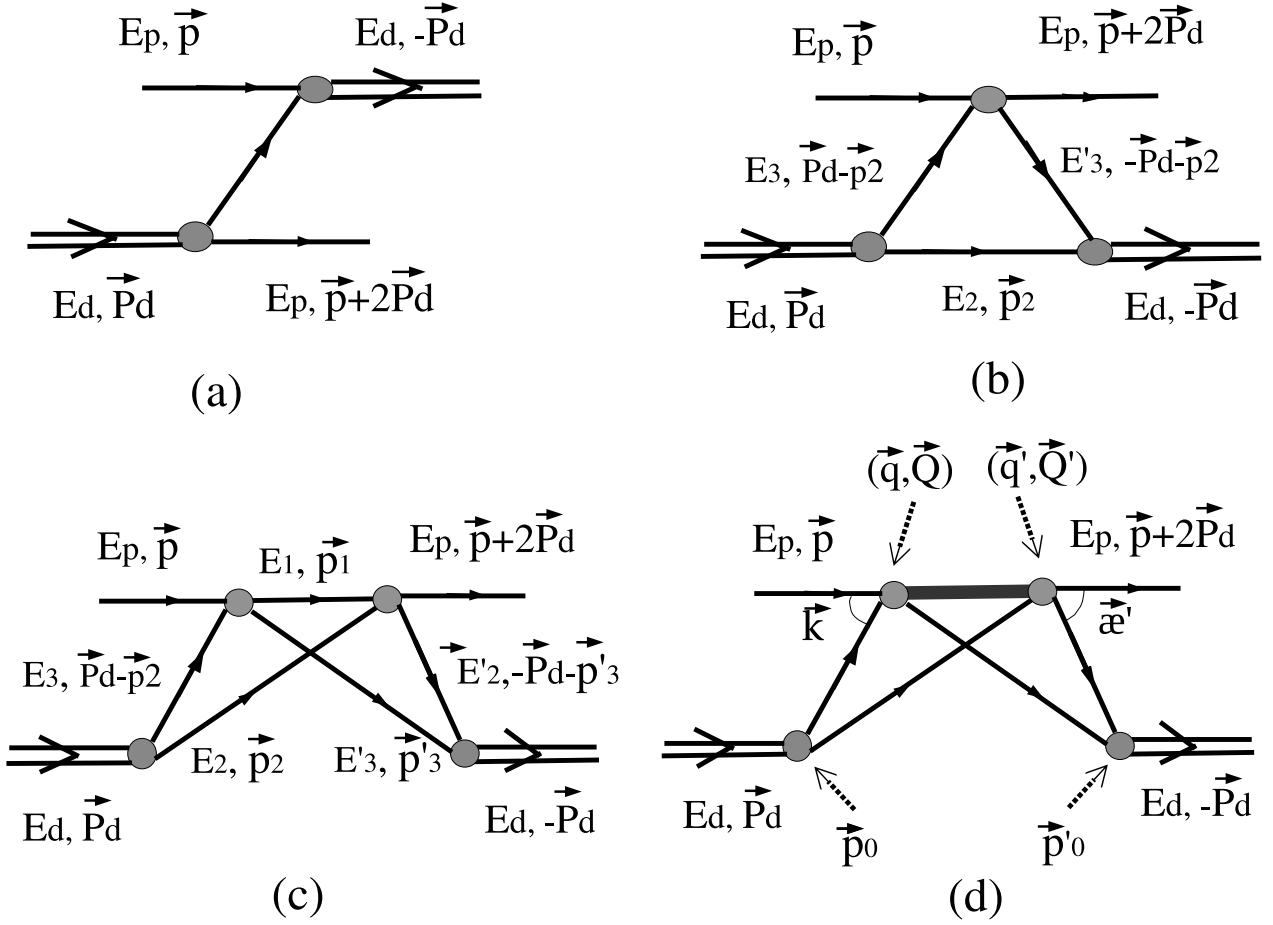}
}
\caption{
The diagrams included into consideration:
(a) the one nucleon exchange diagram; (b) the single
scattering diagram; (c)  the double scattering diagram
with a nucleon in the intermediate state;
(d)the double scattering diagram
with $\Delta$-isobar in the intermediate state.}
\label{diagram}
\end{figure*}
\begin{eqnarray}
\label{e2e3}
&&E_2=\sqrt{m_N^2+\vec p_2^2}~,
~~~~~~
E_3=\sqrt{m_N^2+(\vec P_d -\vec p_2)^2}~,
\nonumber\\
&&E_3^\prime =\sqrt{m_N^2+(\vec P_d +\vec p_2)^2}
\end{eqnarray}
and these energies in the centre-of-mass of the two nucleons
forming the initial and final deuterons
 are equal, correspondingly, to
\begin{eqnarray}
&&E^*=\frac{1}{2}\sqrt{(E_2+E_3)^2-\vec P_d^2},
\\
&&E^{\prime *}=\frac{1}{2}\sqrt {(E_2+E_3^\prime )^2-\vec P_d^2}~~.
\nonumber
\end{eqnarray}

The nucleon-nucleon scattering is described by the t-matrix element. 
 We use the parameterization  of this
 matrix offered by Love and Franey  \cite {LF}. This is the on-shell NN t-matrix
  defined in the cente-of-mass:
\begin{eqnarray}
\label{tnn}
&&<\vec\varkappa^{*\prime}  \mu_1^\prime \mu_2^\prime |t_{c.m.}|
\vec\varkappa^* \mu_1\mu_2>
=<\vec\varkappa^{*\prime}  \mu_1^\prime \mu_2^\prime |
A+
\nonumber\\
&&B(\mbox{\boldmath$\sigma_1$} \hat N^*)(\mbox{\boldmath$\sigma_2$} \hat N^*)+
C(\mbox{\boldmath$\sigma_1$} +\mbox{\boldmath$\sigma_2$} )\cdot \hat N^* +
\\
&&D(\mbox{\boldmath$\sigma_1$} \hat q^*)(\mbox{\boldmath$\sigma_2$} \hat q^*) +F(\mbox{\boldmath$\sigma_1$} \hat Q^*)(\mbox{\boldmath$\sigma_2$} \hat Q^*)
|\vec\varkappa^* \mu_1\mu_2>.
\nonumber
\end{eqnarray}
The orthonormal basis $\{\hat q^*,\hat Q^*,\hat N^*\}$ is a combination of the nucleon
relative momenta in the initial $\vec\varkappa ^*$ and final 
$\vec\varkappa^{\prime *}$
states:
\begin{eqnarray}
\hat q^*=\frac {\vec\varkappa^* -\vec\varkappa^{*\prime}}
{|\vec\varkappa^* -\vec\varkappa^{*\prime}|},~~
\hat Q^*=\frac {\vec\varkappa^* +\vec\varkappa^{*\prime} }
{|\vec\varkappa^* +\vec\varkappa^{*\prime}|},~~
\hat N^*=\frac {\vec\varkappa^*  \times 
\vec\varkappa^{*\prime} }{|\vec\varkappa^*
\times\vec\varkappa^{*\prime} |}.
\nonumber
\end{eqnarray}
The amplitudes $A,B,C,D,F$ are the functions of the centre-of-mass
energy and scattering angle. The radial parts of these amplitudes are taken 
as a sum of Yukawa terms. A new fit of the model parameters \cite{newlf} was done  
in accordance with the  phase-shift-analysis data SP07 \cite{said}.

Since the matrix elements are expressed
via the effective $NN$-interaction operators sandwiched
between the initial and final plane-wave states, this construction 
can be extended to the off-shell case allowing the initial and final
states to get the current values of   $\vec\varkappa$ and
$\vec\varkappa^\prime$. Obviously, this extrapolation does 
not change the general spin structure.

The double scattering term can be divided into  two parts: rescattering with a
nucleon and $\Delta$-isobar in the intermediate state.  
\begin{eqnarray}
&&{\cal J}_{DS}+{\cal J}_{\Delta} =2<N(1)|<\psi_d(23)|N(2)N(3)>
\nonumber\\
&&<N(2)N(3)|[1-P_{12}]|t_3|\{N(1)N(2)+\Delta(1)N(2)\}>
\nonumber\\
&&g_0<\{N(1)N(2)+\Delta(1)N(2)\}|t_2[1-P_{13}]|
\\
&&<N(2)N(3)\psi_d(23)>|N(1)>
\nonumber
\end{eqnarray}
%\begin{eqnarray}
%{\cal J}_{DS}&=&2<1(23)|t^{sym}_3(NN) g_0 t^{sym}_2(NN)|1(23)>~~,
%\end{eqnarray}

The double scattering contribution with the intermediate nucleon (DS) is defined
 by a deuteron wave function and two
nucleon-nucleon t-matrixes. Also we have, here, three-nucleon propagator:  
\begin{eqnarray}
\label{ds}
&&{\cal J}_{DS}=\int d\vec{p_2}d\vec{p_3^\prime}
<-\vec P_d {\cal M}_d^\prime|\Omega_d^\dagger 
\nonumber
\\
&&|-\vec P_d-\vec p_3^\prime~m_2^\prime,
\vec p_3^\prime~ m_3^\prime>
\nonumber
\\
&&<\vec p^\prime~ m^\prime,
 -\vec P_d-\vec p_3^\prime~m_2^\prime, \vec p_3^\prime~m_3^\prime|
\{ 
t_{3(NN)}^1(E^\prime) 
t_{2(NN)}^1(E)+
\nonumber\\
&&[t_{3(NN)}^1(E^\prime)+
t_{3(NN)}^0(E^\prime)]
[t_{2(NN)}^1(E)+
t_{2(NN)}^0(E)]/4\}
\nonumber\\
&&\{E_d+E_p-E_1 -E_2-E_3^\prime +i\varepsilon\}^{-1}
\\
&&|\vec p~m, \vec p_2~ m_2, \vec P_d -\vec p_2~m_3>
%&&
%\frac{1}{
%}
\nonumber\\
&&<\vec p_2~ m_2,~\vec P_d -\vec p_2~ m_3|\Omega_d|\vec P_d {\cal M}_d>.
\nonumber
\end{eqnarray}
The argument of the $NN$-matrix is defined
as the three-nucleon on-shell energy excluding the energy of the nucleon which does not
participate in the interaction:
\begin{eqnarray}
E=E_d+E_p-E_2,~~~~~~E^\prime=E_d+E_p-E^\prime _3.
\end{eqnarray}

The superscript at the t-matrix in Eqs.(\ref{ss}), (\ref{ds}) refers to the isotopic momentum of the nucleon-nucleon pair.
\begin{figure*}[t]
%\resizebox{0.5\textwidth}{!}{
\includegraphics{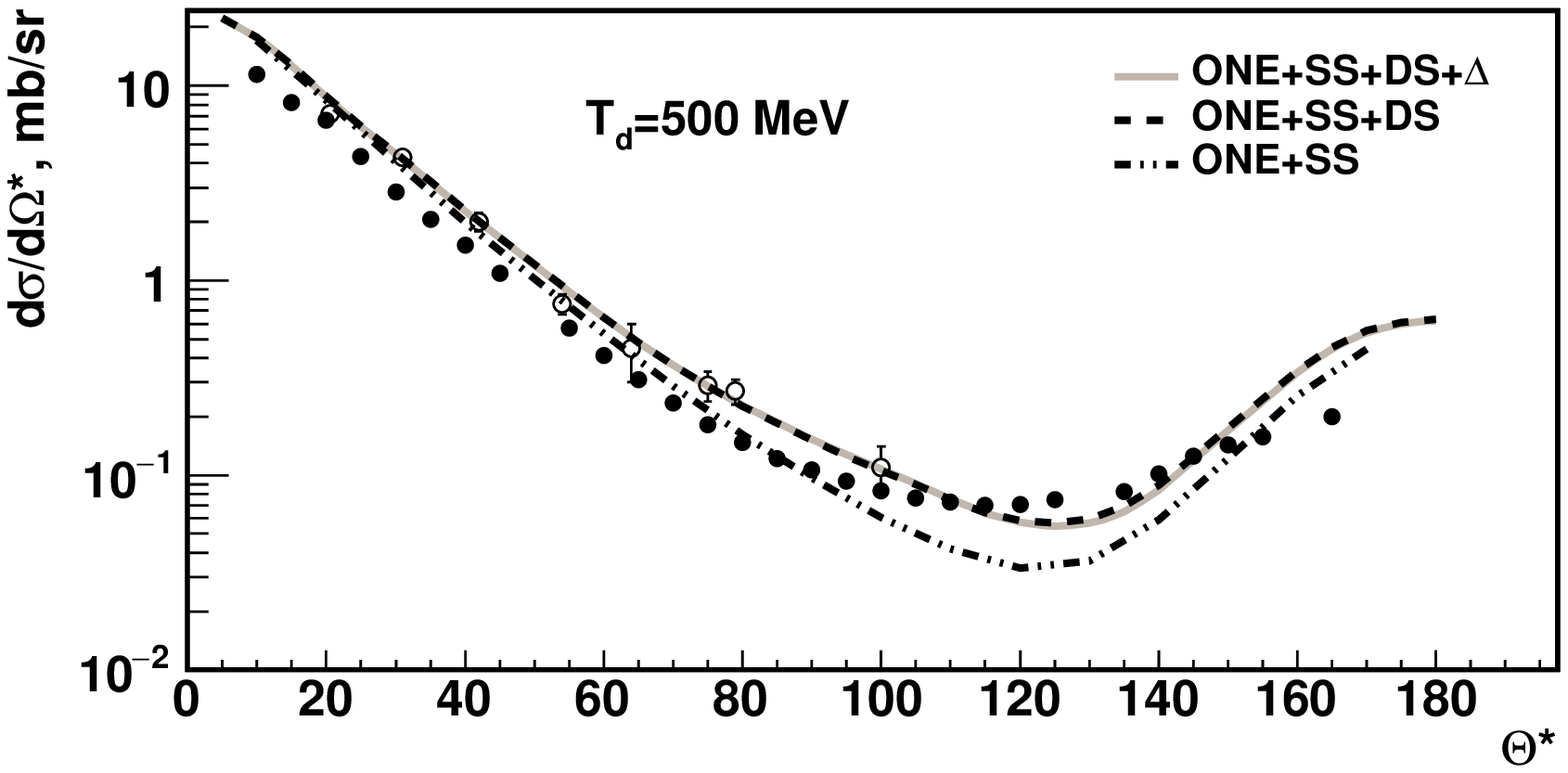}
%}
\caption{The differential cross section at the deuteron kinetic 
energy of 500 MeV as a function of the c.m. scattering angle.
The data are taken from\cite{cr500} ($\bullet$), and\cite{sham}($\circ$).}
\label{f500}
\end{figure*}

\section{$\Delta$-isobar contribution.}

The structure of the amplitude with $\Delta$ in the intermediate state  looks like the double-scattering one. But here we have
$NN\to \Delta N$ matrixes instead of the nucleon-nucleon matrixes and $NN\Delta$-propagator instead of the three-nucleon one.
%The delta amplitude can be written as
\begin{eqnarray}
\label{delta}
&&{\cal J}_{\Delta}=2\int d\vec{p_2}d\vec{p_3^\prime}dE_\Delta d\vec{p_\Delta}\delta(E_\Delta-\sqrt{\mu^2+\vec{p_\Delta^2}})
\nonumber\\
&&\delta(\vec p+\vec {P_d}-\vec{p_2}-\vec{p_3}-\vec{p_\Delta})
\nonumber\\
&&_1<\frac{1}{2}\tau^\prime\frac{1}{2} m^\prime\vec{p^\prime}|_{23}<00;\vec{-P_d}1{\cal M}_d^\prime |\Omega_d^\dagger[1-P_{12}]|
t_{3(N\Delta)}(E^\prime)
\nonumber\\
&&\frac{1}{E-E_2-E_3^\prime-E_\Delta-i\Gamma(E_\Delta/2)}|
\Psi_{\vec p_\Delta}(E_\Delta)>_1
\nonumber\\
&&|\frac{1}{2}\tau_2\frac{1}{2}m_2\vec p_2;
\frac{1}{2}\tau_3\frac{1}{2}m_3\vec p_3>_{23}
\\
&& _{23}<\frac{1}{2}\tau_2\frac{1}{2}m_2\vec p_2;
\frac{1}{2}\tau_3\frac{1}{2}m_3\vec p_3| _1<\Psi_{\vec p_\Delta}(E_\Delta)|
\nonumber\\
&&t_{2(N\Delta)}(E)[1-P_{13}]\Omega_d|
\vec{P_d} 1{\cal M}_d; 00>_{23}|\frac{1}{2}\tau\frac{1}{2} m\vec{p}>_1
\nonumber
\end{eqnarray}

Moreover, the $\Delta$-isobar does not have a fixed mass and is defined by the mass-distribution.
Therefore, we save here the integration over the $\Delta$-energy, $E_\Delta$.
A full set of the particles quantum numbers was included into the amplitude definition in Eq.(\ref{delta}).
Isospin and spin quantum numbers are marked by $\tau$ and $m$ or ${\cal M}_d$, respectively. The indexes
near the bracket correspond to the particles numbers. 
%\begin{eqnarray}
%E_i=\sqrt{m^2+\vec p_i^2}
%\\
%E\equiv E_{on-shell}=\sqrt{M_d^2+\vec P_d^2}+\sqrt{m^2+\vec p^2}
%\nonumber
%\end{eqnarray}

The distribution function of the $\Delta$-energy
\begin{eqnarray}
|\Psi_{\vec p_\Delta}(E_\Delta)><\Psi_{\vec p_\Delta}(E_\Delta)|=\rho(E_\Delta)
\end{eqnarray}
is defined through the $\Delta$-width $\Gamma(\mu)$:
\begin{eqnarray}
\rho(\mu)=\frac{1}{2\pi}\frac{\Gamma(\mu)}{(E_\Delta (\mu)-E_\Delta(m_\Delta))^2+\Gamma^2(\mu)/4},
\end{eqnarray}
where $\mu^2=E_\Delta^2-\vec p^2_\Delta$ is a squared current mass of the delta.
The delta width is energy dependent. We use, here, a standard parameterization of $\Gamma(\mu)$ taking into account
the $\Delta$ off-shell corrections:
\begin{eqnarray}
\Gamma(\mu)=\Gamma_0\frac{p^3(\mu^2,m_\pi^2)}{p^3(m^2_\Delta,m_\pi^2)}\cdot\frac{p^2(m^2_\Delta,m_\pi^2)+\gamma^2}{p^2(\mu^2,m_\pi^2)+\gamma^2}.
\end{eqnarray}
where $p(x^2,m_\pi^2)$ is the momentum value in the $\pi N$ -center-of-mass:
\begin{eqnarray}
p(x^2,m_\pi^2)=\sqrt{(x^2+m_N^2-m_\pi^2)^2/4x^2-m_N^2}.
\end{eqnarray}
All parameters were taken from ref.\cite{jain}:
%In our calculation we use the following value:
\begin{eqnarray} 
\Gamma_0=0.120 ~GeV,~~\gamma=0.200 ~GeV,~~m_\Delta=1.232 
\end{eqnarray}
%\begin{eqnarray}
%\mu^2=E_\Delta^2-\vec p^2_\Delta
%\\
%dE_\Delta=\frac{\mu}{E_\Delta}d\mu
%\end{eqnarray}

In the  Born approximation the $NN\to N\Delta$ t-matrix  can be replaced with the
corresponding potential $V_{(N\Delta)}(E)$:
\begin{eqnarray}
&&<\vec p,\frac{1}{2}m,\frac{1}{2}\tau|t_{(N\Delta)}(E)|\Psi_{\vec p_\Delta}(E_\Delta)>\approx
\nonumber\\
&&<\vec p,\frac{1}{2}m,\frac{1}{2}\tau|V_{(N\Delta)}(E)|\Psi_{\vec p_\Delta}(E_\Delta)>
\end{eqnarray}

The potential for the $NN\to N\Delta$ transition is based on  the $\pi-$ and $\rho-$ exchanges:
\begin{eqnarray}
V_{\beta\alpha}^{(\pi)}&=&-\frac{f_\pi f_\pi^*}{m_\pi^2}F_\pi^2(t)\frac{q^2}{m_\pi^2-t}
(\vec\sigma\cdot\hat q)(\vec S\cdot \hat q)(\vec\tau\cdot\vec T)
\\
V_{\beta\alpha}^{(\rho)}&=&-\frac{f_\rho f_\rho^*}{m_\rho^2}F_\rho^2(t)\frac{q^2}{m_\rho^2-t}
\{(\vec\sigma\vec S)-(\vec\sigma\cdot\hat q)(\vec S\cdot \hat q)\}(\vec\tau\cdot\vec T)
\nonumber
\end{eqnarray}
Here, $t$ is a four-momentum transfer and $\vec q$ is a corresponding three-momentum transfer.
Operators $\vec\sigma (\vec \tau)$  are $\frac{1}{2}$- spin (isospin) operators defined by Pauli
matrixes while $\vec S (\vec T)$ operators correspond to $\frac{1}{2}\to \frac{3}{2}$ spin (isospin) transition.
$m_\pi$ and $m_\rho$ are  pion and $\rho$- meson masses. 
The coupling constant $f_\pi$ is related with the $NN\pi$ vertex and $f_\pi^*$ corresponds to the 
$N\Delta\pi$ one. It  also concerns $\rho-$ coupling constants.
\begin{eqnarray}
f_\pi&=&1.008~~~~~~f_\pi^*=2.156
\\
f_\rho&=&7.8~~~~~~f_\rho^*=1.85f_\rho
\end{eqnarray}
The hadronic form factor was chosen in the monopole form:  
\begin{eqnarray}
F_x(t)=(\Lambda_x^2-m_x^2)/(\Lambda_x^2-t)
\end{eqnarray}
In our calculation we use $\Lambda=1$  GeV. 

The presented approach for the  $\Delta$-isobar term was applied in refs. \cite{jain}, \cite{dmitr}  to describe
the data on the $\Delta$- production in the
$pp\to n\Delta^{++}$.
 The obtained results were in a good agreement with the experimental data on the differential
cross section.
%, what tell us about a correctness this model.
%All parameters were taken from ref.\cite{jain}. 
%The correctness of the such  description of
%the $\Delta$-excitation was demonstrated in \cite{jain}, \cite{dmitr}, where the data on the $\Delta$-isobar production in the
%$pp\to n\Delta$ were analysed.

Since two nucleon states in the $NN\to N\Delta$ vertexes are antisymmetrized, two permutation operators appear in Eq.(\ref {delta}).
As consequence, the $\Delta$- amplitude contains four terms: one direct, two exchange, and one double-exchange ones.  
The permutation operator $P_{ij}$ involves the permutation of all quantum numbers. Here, it is permutation over momentum, spin, and
isospin indexes:   $P_{ij}=P_{ij}(p)P_{ij}(\sigma)P_{ij}(\tau)$.

The isotopic coefficient in Eq.(\ref{delta}) is equal for the all four contributions into the $\Delta$- amplitude and defined as 
\begin{eqnarray}
c_T=<\frac{1}{2}\tau_2^\prime\frac{1}{2}\tau_3^\prime|00><\frac{1}{2}\tau^\prime\frac{1}{2}\tau_2^\prime|(\vec T_1 \vec\tau_2)|
\frac{3}{2}\tau_\Delta\frac{1}{2}\tau_2>
\\
<\frac{3}{2}\tau_\Delta\frac{1}{2}\tau_3^\prime|(\vec T_1 \vec\tau_3)|
\frac{1}{2}\tau\frac{1}{2}\tau_3><\frac{1}{2}\tau_2\frac{1}{2}\tau_3|00>.
\nonumber
\end{eqnarray}
It can be calculated independently on the spin part of the $\Delta$- amplitude:
\begin{eqnarray}
c_T=\frac{1}{12}<\frac{1}{2}\|T\|\frac{3}{2}>^2<\frac{1}{2}\|\tau\|\frac{1}{2}>^2=2,
\nonumber
\end{eqnarray}
where the reduced matrix elements  $<\frac{3}{2}\|T\|\frac{1}{2}> =$ \\
$ <\frac{3}{2}\|S\|\frac{1}{2}> = 2$ and
$<\frac{1}{2}\|\tau\|\frac{1}{2}> = <\frac{1}{2}\|\sigma\|\frac{1}{2}> = \sqrt{6}$ were substituted.
%\begin{eqnarray}
%
%\end{eqnarray}

The permutation over spin states is defined in the same way it was done for the one-nucleon exchange term (\ref{ONE}). 
\begin{figure*}
\includegraphics{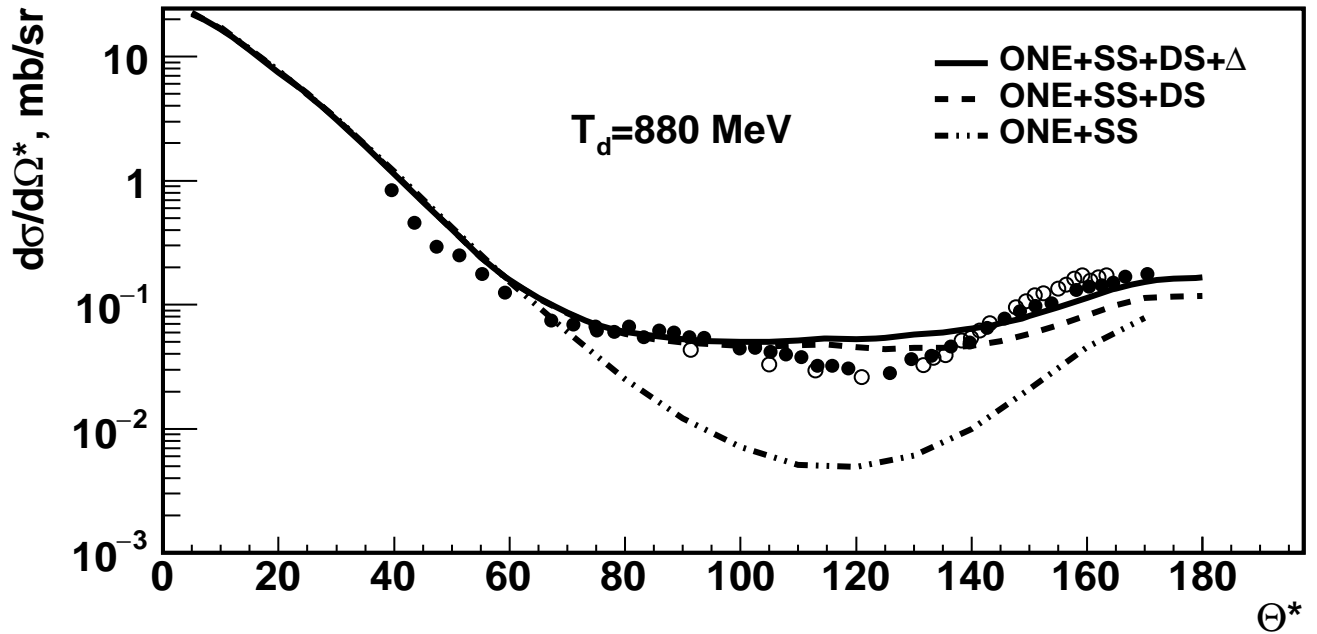}
\caption{The differential cross section at the deuteron kinetic 
energy of 880 MeV as a function of the c.m. scattering angle.
The data are taken from \cite{cr880_a} ($\bullet$), and\cite{alder}($\circ$).}
\label{f880}
\end{figure*}
\subsection{Kinematics}

We consider, here,  the kinematics of the double scattering diagram in detail.
All the presented  expressions concern the diagram  with a $\Delta$- excitation.
 However, they are also correct for the double-scattering diagram with a nucleon 
 in the intermediate state, if $(E_\Delta,\vec p_\Delta)$ is replaced with $(E_1,\vec p_1)$.
 The momenta notations are shown in Fig.(1c).

The two nucleons relative momenta in the deuterons vertexes are defined in relativistic
kinematics in the following manner:
\begin{eqnarray}
\vec{p_0}&=&\frac{(E_2+E^*)\vec p_3-(E_3+E^*)\vec p_2}{E_2+E_3+2E^*}
\\
\vec{p_0^\prime}&=&\frac{(E_2^\prime+E^{\prime *})\vec p^\prime_3-(E_3^\prime+E^{\prime *})\vec p^\prime_2}
{E^\prime_2+E^\prime_3+2E^{\prime *}}.
\nonumber
\end{eqnarray}
Here, $E^*$ and $E^{\prime *}$ are the energies one of the nucleon in the corresponding deuteron rest frame,
which is equivalent the two nucleons centre-of-mass:
\begin{eqnarray}
E^*=\sqrt{(p_2+p_3)^2}/2,~~~~~E^{\prime *}=\sqrt{(p_2^\prime+p_3^\prime)^2}/2
\end{eqnarray}

In accordance to the momentum conservation in the deuteron vertexes we can get the following
expressions for the internal deuteron momenta in the deuterons Breit frame:
\begin{eqnarray}
\vec{p_0}&=&\frac{\vec P_d}{2}-\vec p_2-\frac{E_2-E_3}{E_2+E_3+2E^*}\frac{\vec P_d}{2}
\\
\vec{p_0^\prime}&=&\frac{\vec P_d}{2}+\vec p_3^\prime+\frac{E_2^\prime-E_3^\prime}
{E^\prime_2+E^\prime_3+2E^{\prime *}}\frac{\vec P_d}{2}
\nonumber
\end{eqnarray}

Note, that two first terms in these equations correspond to the ordinary definition of the nonrelativistic
relative momentum. 

The transfer momenta in the $NN\to\Delta N$ vertexes, $\vec q$ and $\vec q^\prime$, are defined in a standard manner:
\begin{eqnarray}
\vec q=\frac{1}{2}(\vec k^\prime -\vec k)~~~~ \rm{and}~~~\vec q^\prime=\frac{1}{2}(\mbox{\boldmath$\varkappa$}^\prime -\mbox{\boldmath$\varkappa$}), 
\end{eqnarray}
where vectors $ \vec k, \vec k^\prime$ are the nucleon-nucleon and nucleon-delta relative momenta
in the $NN\to\Delta N$ vertex :
\begin{eqnarray}
\vec{k}&=&\frac{(E_3+E^*_{13})\vec p-(E_p+E^*_{13})\vec p_3}{E_p+E_3+2E^*_{13}}
\\
\vec{k^\prime}&=&\frac{(E_3^\prime+E^{\prime *}_{13})\vec p_\Delta-(E_\Delta+E^{\prime *}_{13})\vec p^\prime_3}
{E_\Delta+E^\prime_3+2E^{\prime *}_{13}}
\nonumber
\end{eqnarray}
and  $\mbox{\boldmath$\varkappa$}, \mbox{\boldmath$\varkappa$}^\prime$ are the nucleon-delta and nucleon-nucleon   relative momenta
in the $\Delta N\to NN$ vertex :
\begin{eqnarray}
\mbox{\boldmath$\varkappa$}&=&\frac{(E_2+E^*_{12})\vec p_\Delta-(E_\Delta+E^*_{12})\vec p_2}{E_\Delta+E_2+2E^*_{12}}
\\
\mbox{\boldmath${\varkappa^\prime}$}&=&\frac{(E_2^\prime+E^{\prime *}_{12})\vec p^\prime-(E_p+E^{\prime *}_{12})\vec p^\prime_2}
{E_p+E^\prime_2+2E^{\prime *}_{12}}.
\nonumber
\end{eqnarray}

$E_{ij}^*=\sqrt{(p_i+p_j)^2}/2$ is the invariant variable defined by $p_i~ {\rm and}~  p_j$ four-momenta.

In this section we considered only a direct term in the  amplitude with the delta in the intermediate state. The other terms can be obtained by the replacement the
corresponding momentum with one with an opposite sign. In such a way $\vec q=(\vec k^\prime -\vec k)/2$, for example, replaces with 
$\vec Q=(\vec k^\prime +\vec k)/2$
etc.
%\newpage
\begin{figure*}[t]
\includegraphics{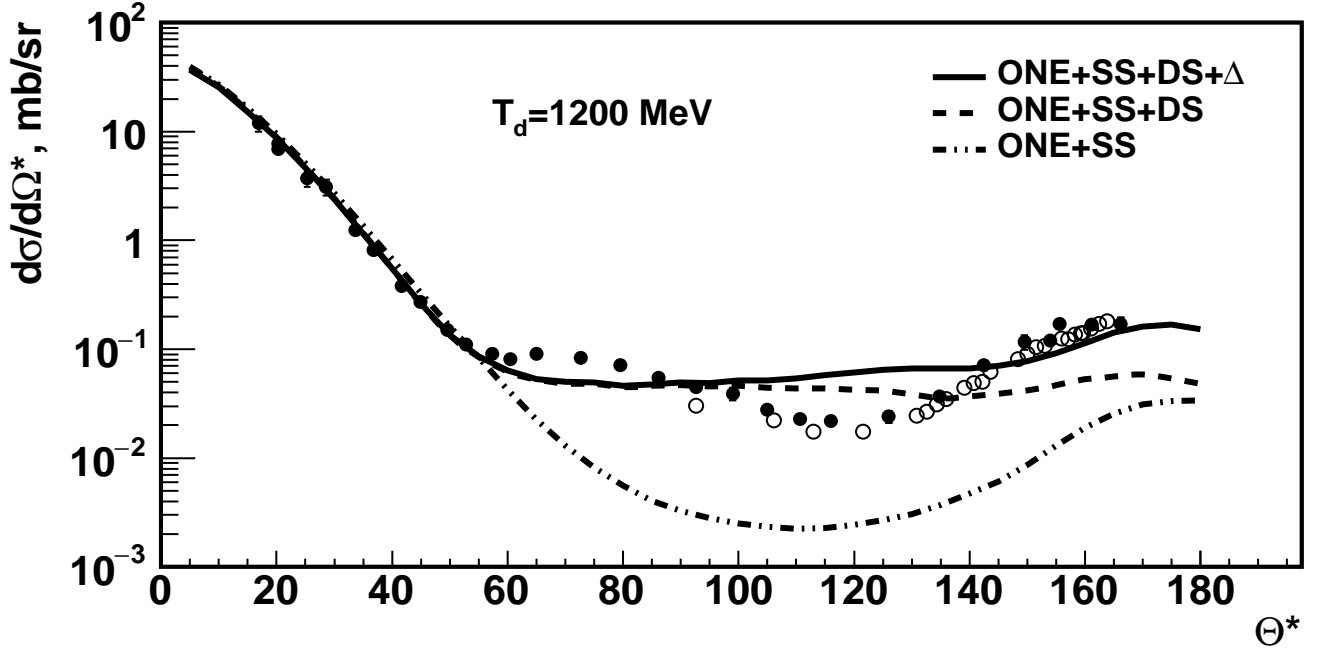}
\caption{The differential cross section at the deuteron kinetic 
energy of 1200 MeV as a function of the c.m. scattering angle.
The data are taken from \cite{cr1200} ($\bullet$), and \cite{alder} ($\circ$).}
\label{f1200}
\end{figure*}

\subsection{Spin part}

To simplify the calculation in this paper we consider the reduced $N\to \Delta$ t-matrix.
The $\rho$-meson part is not taken into account. We believe that the $\pi$-meson term
gives the general contribution into the differential cross section at backward angles.    
However,  $\rho$-meson can  be very important to describe some polarisation observables
due to its vector nature.
Therefore, it will be included into consideration in further calculations. 

In such a way spin structure of the $\Delta$-amplitude is defined by the expression:
\begin{eqnarray}
&&_{(23)}<1{\cal M}^\prime|u(p_0^\prime)+
\nonumber\\
&&\frac{w(p_0^\prime)}{\sqrt{2}}
\left[3(\vec\sigma_2\hat {p_0^\prime})(\vec\sigma_3\hat {p_0^\prime})-(\vec\sigma_2\vec\sigma_3)\right]|
\frac{1}{2}m_2^\prime \frac{1}{2}m_3^\prime>
\nonumber\\
&&<\frac{1}{2}m_2^\prime|(\vec\sigma_2\hat{q^\prime})|\frac{1}{2}m_2><\frac{1}{2}m_3^\prime|(\vec\sigma_3\hat{q})|\frac{1}{2}m_3>
\nonumber\\
&&<\frac{1}{2}m_2\frac{1}{2}m_3|u(p_0)+
\\
&&\frac{w(p_0)}{\sqrt{2}}
\left[3(\vec\sigma_2\hat {p_0})(\vec\sigma_3\hat {p_0})-(\vec\sigma_2\vec\sigma_3)\right]|1{\cal M}>_{(23)}
\nonumber\\
&&<\frac{1}{2}m^\prime|(\vec S_1\hat{q^\prime})|\frac{3}{2}M_\Delta><\frac{3}{2}M_\Delta|(\vec S_1\hat{q})|\frac{1}{2}m>.
\nonumber
\end{eqnarray}

After straightforward  calculations we get some combinations of $\sigma_{1,i},~\sigma_{1,i}\sigma_{2,j}$ and etc. 
sandwiched between deuterons spin states.
It is useful to apply the following  relations:
\begin{eqnarray}
&&_{(23)}<1{\cal M}^\prime|\vec\sigma_{2,i}|1{\cal M}>_{(23)}=
\nonumber\\
&&_{(23)}<1{\cal M}^\prime|\vec\sigma_{3,i}|1{\cal M}>_{(23)}=
\nonumber\\
&&_{(23)}<1{\cal M}^\prime|\vec S_{d,i}|1{\cal M}>_{(23)}
\\
&&_{(23)}<1{\cal M}^\prime|\vec\sigma_{2,i}\vec\sigma_{3,j}|1{\cal M}>_{(23)}>=
\nonumber\\
&&<1{\cal M}^\prime|\frac{1}{3}\delta_{ij}+2Q_{ij}|1{\cal M}>_{(23)},
\nonumber
\end{eqnarray}
where $\vec S_d$ and $Q_{ij}$ are the  spin and    tensor operators of a deuteron, respectively.
Also the following expression  
\begin{eqnarray}
&&<\frac{1}{2}m^\prime|(\vec S\hat{q^\prime} )|\frac{3}{2} M_\Delta><\frac{3}{2} M_\Delta|(\vec S\hat{q} )|\frac{1}{2}m>=
%\nonumber
\nonumber\\
&&\frac{1}{6}<\frac{1}{2}\|S\|\frac{3}{2}>^2[(\hat{q}\hat{q^\prime})<\frac{1}{2}m^\prime|I|\frac{1}{2}m>+
\\
&&\frac{i}{2}<\frac{1}{2}m^\prime|(\vec \sigma\hat{q}\times \hat{q^\prime})|\frac{1}{2}m>]
\nonumber
\end{eqnarray}
is very helpful to simplify the calculations with 3/2-spin operators.
%\newpage
\begin{figure*}[t]
\includegraphics{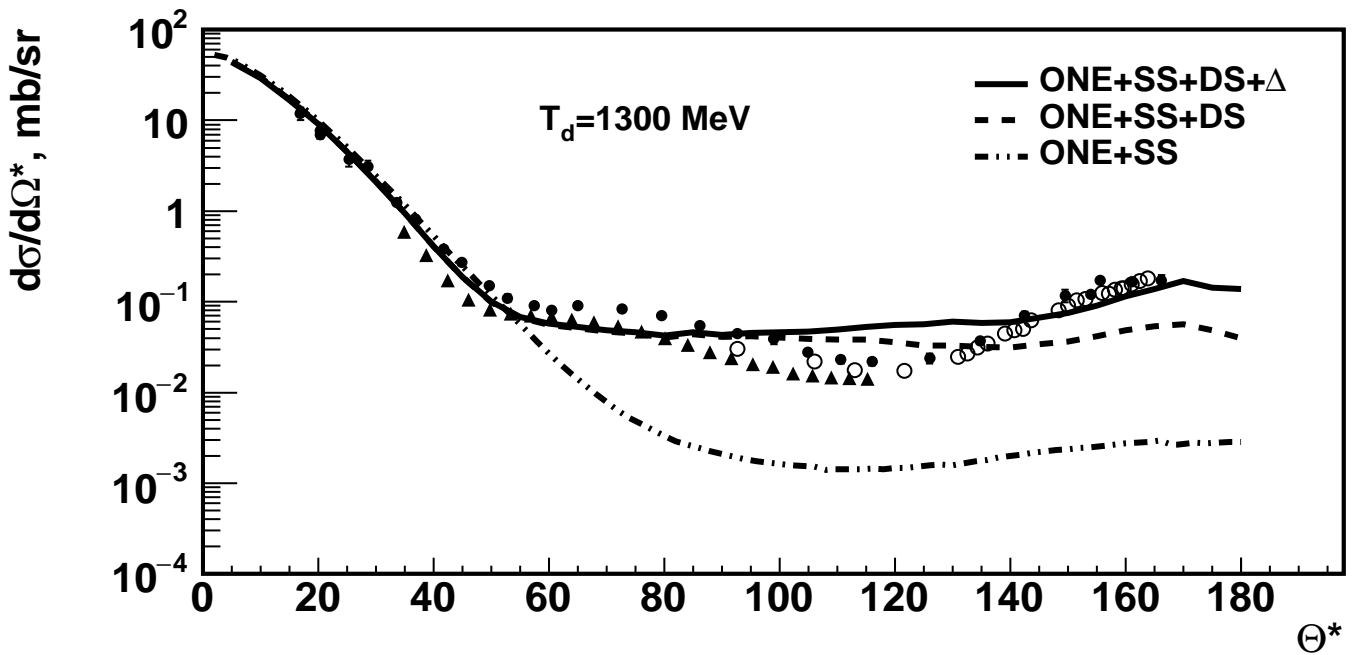}
\caption{The differential cross section at the deuteron kinetic 
energy of 1300 MeV as a function of the c.m. scattering angle.
The data are taken for $T_d=1200$ MeV from \cite{cr1200}($\bullet$), $T_d=1180$ MeV \cite{alder} ($\circ$),
 and $T_d=1282$ MeV \cite{cr1300} ($\blacktriangle$).}
\label{f1300}
\end{figure*}

\section{Results and discussions.}

We applied the presented above model to describe the differential cross sections.
The formal definition of the total cross section of deuteron-proton elastic scattering  is given by
\begin{eqnarray}
&&\sigma(dp\to dp)=(2\pi)^4 \frac{1}{6}\int \frac{d\vec P_d^\prime}{E_d^\prime}\frac{d\vec p^\prime}{E_p^\prime}
\delta(E_d+E_p-E_d^\prime-E_p^\prime)
\nonumber\\
&&\delta(\vec P_d+\vec p-\vec P_d^\prime-\vec p^\prime)
\frac{|T|^2}{\sqrt{(P_d p)^2-P_d^2 p^2}},
\end{eqnarray}
where $|T|^2$ is the squared invariant amplitude defined as
\begin{eqnarray}
 |T|^2=|\sqrt{E_d^\prime E_p^\prime}~{\cal J}_{dp\to dp}~\sqrt{E_d E_p}|^2\equiv inv.
\end{eqnarray}
The expression for the differential cross section
in the center-of-mass can be obtained through the scattering amplitude in the Breit frame as:
\begin{eqnarray}
\frac{d\sigma}{dcos\theta^*}=\frac{(2\pi)^5}{24s}\left(s-m_N^2-M_d^2+\frac{t}{2}\right)^2|{\cal J}_{dp\to dp}^{(Breit)}(s,t)|^2,
\nonumber
\end{eqnarray}
where $s=(P_d+p)^2$ and $t=(p-p^\prime)^2$ are  invariant Mandelstam variables.

We evaluated the dp-elastic differential cross section at four energies in a wide range.
The results presented for three cases: 1) only one-nucleon exchange (ONE) and single-scattering (SS) diagrams taken into account,
2) the double-scattering (DS) term is added into the consideration, and 3) the $\Delta$-isobar excitation is included.

At first, the deuteron kinetic energy equal to 500 MeV was considered (Fig.\ref {f500}).
One can see, the SS-diagram plays an important role up to the scattering angle equal to about $60^\circ$. 
 The DS-contribution is significant at the scattering angle between $60^\circ$ and  $160^\circ$.
The difference between the results taking into account the double scattering and without it reaches 2-3 times at $\theta^*=120^\circ$.
  
This energy is not high enough for  the $\Delta$-isobar manifestation.  Therefore, the inclusion of the $\Delta$-contribution
term into the scattering amplitude does not have to influence remarkably to the result. In fact, the results obtained
with and without the  $\Delta$-isobar term are undistinguished in a whole angular range. 

The second energy is equal to 880 MeV (Fig.\ref {f880}). One can see, the contribution of the double-scattering
is very significant at this energy. The inclusion of the DS-term into consideration allows to describe the behaviour of the 
differential cross section at the scattering angle range  $60^\circ - 140^\circ$. The difference between ONE+SS curve and ONE + SS + DS
one reaches about 15 times. Note, the  $\Delta$ excitation begins to manifest itself at the angle equal to about $120^\circ $ and describes 
the behaviour of the experimental data at the angle above $140^\circ $ where the differential cross section sharply increases.

The dp-elastic differential cross sections at energies equal to  $T_d=1200$  MeV and $T_d=1300$  MeV are presented  in
Fig.\ref {f1200} and Fig.\ref {f1300}, respectively. Here we can observe further enhancement of the double scattering term influence
on the result.
Also the contribution of the  $\Delta$ - isobar increases significantly at the scattering angle above $140^\circ $ .

Analysing the obtained results one can conclude the following. The differential cross sections are described quite well up to
the scattering angles equal to
 $60^\circ$ taking into account only the one-nucleon-exchange and single-scattering terms.
If we consider the dp-elastic scattering at the angles large than  $60^\circ$,
 it is necessary to include the
double scattering term into consideration.
It should be noted 
 the double-scattering contribution into the reaction
amplitude increases with the deuteron energy growing and may change the value of the differential cross section
 on a few orders in comparison with the result obtained without  inclusion of the  DS-term. 

 The sharp rise of the differential cross
section at  $\theta^*\ge140^\circ $, which is observed in the experimental data,
points at the appearance of the addition reaction mechanism. The inclusion of the  
$\Delta$-excitation term into consideration allowed to describe this rise of the 
differential cross section at the large scattering angles.
The contribution  of the  $\Delta$-isobar mechanism grows with the initial deuteron energy.
It is negligible at $T_d=500$ MeV and significant at    $T_d=1300$ MeV.

In such  way we got a good description  of the differential cross section of 
deuteron-proton elastic scattering in a whole angular range, from $0^\circ $ to  $180^\circ $ of the scattering angle,
in a wide deuteron energy interval, between 500 MeV and 1300 MeV.

 \vspace{1cm}
{\it Acknowledgements:}
The author is grateful to Dr. V.P. Ladygin for fruitful discussions and interest in this problem.
This work has been supported by the Russian Foundation for Basic Research
under grant  $N^{\underline 0}$ No.16-02-00203a.

%
%\newpage
%

%


\begin{thebibliography}{99}

\bibitem{sham} R.D.Shamberger, {\it Phys.Rev.\/} {\textbf 85}, 424 (1952). 

\bibitem {cham} O.Chamberlain, M.O.Stern, Phys.Rev.\textbf{94}, (1954) 666

\bibitem {clark} O.Chamberlain, D.D.Clark, Phys.Rev.\textbf{102}, (1956) 473

\bibitem {crewe} A.V.Crewe, et al., Phys.Rev.\textbf{114}, (1959) 1361

\bibitem {marshal} L.Marshall, et al., Phys.Rev.\textbf{95}, (1954) 1020

\bibitem {marc} S.Marcowitz, Phys.Rev.\textbf{120}, (1960) 891

\bibitem {postma} H.Postma, R.Wilson, Phys.Rev.\textbf{121}, (1961) 1229 

\bibitem {igo1} G. Igo et al., Nucl.Phys.,\textbf{A195}, (1972) 33,

\bibitem {igo2} M.A. Nasser et al., Nucl.Phys., \textbf{A229}, (1974) 113

\bibitem {395} M.Garcon { et al.}, Nucl.Phys.A {\bf 458}, 287 (1986).

\bibitem {1200} J.Arvieux { et al.}, Nucl.Phys.A {\bf 431}, 613 (1984).

\bibitem {1200a} M.Haji-Saied { et al.}, Phys. Rev.C {\bf 36}, 2010 (1987).		  

\bibitem {trpol} Sun Tsu-hsun { et al.} Phys.Rev.C {\bf 31} 515 (1985).

\bibitem {trpol1} A.Rahbar { et al.} Rhys. Lett.B {\bf 194} 338 (1987).

\bibitem {ghaz} V.Ghazikhanian { et al.} Phys.Rev.C {\bf 43} 1532 (1991).


\bibitem{cr500} K.Hatanaka et al., Phys.Rev.\textbf{C66}, (2002) 044002 

\bibitem{maeda} Y.Maeda et al., Phys.Rev.\textbf{C}, (2007) 014004

\bibitem{ermish} K.Ermish et al., Phys.Rev.\textbf{C71}, (2005) 064004
	
\bibitem{glag} V.V. Glagolev, V.P. Ladygin, N.B. Ladygina, A.A. Terekhin,
 Eur.Phys.J. \textbf{A48}, (2012) 182

\bibitem{pkur} P.K.Kurilkin, et al., Phys.Lett.B \textbf{715}, (2012) 61

\bibitem{japh} N.B.Ladygina, Phys.Atom.Nucl.\textbf{71}, (2008) 2039 .

\bibitem{epja} N.B.Ladygina, Eur. J. Phys. A {\bf 42}, (2009)  91.

\bibitem{ker} A.K.Kerman, L.S.Kisslinger, Phys.Rev. {\bf 180}, 1483 (1969).

\bibitem{kon26} L.A.Kondratyuk, F.M.Lev, Sov.J.Nucl.Phys. {\bf26}, 153 (1977). 

\bibitem{kon29} L.A.Kondratyuk, F.M.Lev, L.V.Shevchenko, Sov.J.Nucl.Phys. {\bf29}, 558 (1979). 

\bibitem{kon33} L.A.Kondratyuk, F.M.Lev, L.V.Shevchenko, Yad.Fiz. {\bf33}, 1208 (1981).
 
\bibitem{chiev} A.Deltuva, K.Chmielewski, P.U.Sauer, Phys.Rev.\textbf{C67},
 034001 (2003).

\bibitem{deltuva} A.Deltuva, R.Machleidt, P.U.Sauer, Phys.Rev.\textbf{C68},
  024005 (2003).

\bibitem {ags} E.O.Alt, P.Grassberger, W.Sandhas, 
Nucl.Phys. {\bf B2}, 167 (1967)

\bibitem {zig} E.Schmid, H.Ziegelmann,{\it The Quantum Mechanical Three-Body
Problem} (Oxford, Pergamon Press, 1974).

\bibitem {B} R.Machleidt, K.Holinde, Ch.Elster,
Phys.Rep. \textbf {149}, 1 (1987).

\bibitem {cd} R.Machleidt,  Phys. Rev. \textbf{C63}, 024001 (2001).

\bibitem {par} M. Lacombe et al., Phys.Lett.\textbf{B101}, 139 (1981).

\bibitem{LF} 
 W.G.Love, M.A.Franey, Phys. Rev.C {\textbf 24}, 1073 (1981)


\bibitem{newlf}  N.B.Ladygina (2008), e-preprint nucl-th/0805.3021

\bibitem  {said}
 http://gwdac.phys.gwu.edu

\bibitem{jain} B.K.Jain, A.B.Santra, Phys.Rep.{\textbf 230}, 1 (1993)

\bibitem{dmitr} V.Dmitriev, O.Sushkov, C.Gaarde, Nucl.Phys. {\textbf A459}, 503 (1986)

\bibitem{cr880_a} N.E.Booth {\it et al.}, {\it Phys.Rev.D\/} {\textbf 4},  1261  (1971).

\bibitem{alder}  J.C.Alder {\it et al.}, {\it Phys.Rev.C\/} {\textbf 6},  2010 (1972).

\bibitem{cr1200} E.T.Boschitz {\it et al.}, {\it Phys.Rev.C\/} {\textbf 6},  457 (1972).


\bibitem{cr1300} E.G\"{u}lmez et al.,Phys.Rev.C \textbf{42}, 2067 (1991).
\end{thebibliography}
\end{document}